\documentstyle[preprint,eqsecnum,aps,epsf]{revtex}

\begin{document}
\draft
\preprint{\vbox{
\hbox{INPP-UVA-00-02}
\hbox{April 15, 2000}
\hbox{hep-ph/0008104}
}}

\title{\bf Modified Wandzura-Wilczek Relation\\
with the Nachtmann Variable\\}

\author{Xiaotong Song$^*$ }

\address{Institute of Nuclear and Particle Physics\\ 
Department of Physics, University of Virginia\\ 
P.O.Box 400714, Charlottesville, VA 22904-4714\\}

\maketitle
\begin{abstract}

If one retains $M^2/Q^2$ terms in the kinematics, the Nachtmann
variable $\xi$ seems to be more appropriate to describe deep
inelastic lepton-nucleon scattering. Up to the first power of 
$M^2/Q^2$, a modified Wandzura-Wilczek relation with respect to 
$\xi$ was derived. Kinematical correction factors are given as 
functions of $\xi$ and $Q^2$. A comparison of the modified 
$g_2^{WW}(\xi)$, and original $g_2^{WW}(x)$ with the most 
recent $g_2$ data is shown.  
\end{abstract}
\bigskip
\bigskip
\pacs{13.85.Hd, 13.88.+e, 14.20.Dh\\
\\
$^*$~Email address: xs3e@virginia.edu}

\widetext

The standard Bjorken variable $x_B=Q^2/2P\cdot q$ is commonly used 
in the discussion of deep inelastic scattering (DIS). However, 
if one retains $M^2/Q^2$ terms in the kinematics another variable 
$\xi$ introduced by Nachtmann\cite{nacht} (cf. Greenberg and Bhaumik
\cite{green}) 
$$\xi={2x}/(1+\sqrt {1+4M^2x^2/Q^2}~)
\eqno (1)$$ 
seems to be more appropriate to describe DIS processes. 

For the deep inelastic polarized lepton-nucleon scattering, the asymmetry 
depends on two spin structure functions $g_1$ and $g_2$. The structure 
function $g_1(x, Q^2)$ can be interpreted as a charge-square weighted
quark helicity distribution in the parton model, and the EMC \cite{emc} 
measurement led to a surprising result - the quark spins contribute a very
small fraction of the spin of the proton - the so-called 
`` spin puzzle ''. Since then many theoretical works and experimental 
measurements have been done to solve this `` puzzle '' \cite{review}. 
Most recently, the second spin structure function $g_2$, which includes
both twist-2 and twist-3 contributions, has been measured with 
relatively higher precision \cite{e155}. Using the operator product
expansion (OPE) approach, one can obtain a relation between $g_1$ and
$g_2$
$$g_2(x,Q^2)=g_2^{WW}(x,Q^2)+\bar g_2(x,Q^2),
\eqno (2)$$
where
$$g_2^{WW}(x,Q^2)\equiv -g_1(x,Q^2)+\int_{x}^{1}{{dy}\over y}g_1(y,Q^2).
\eqno (3)$$
If the twist-3 contribution $\bar g_2(x,Q^2)$ can be neglected, then 
Eq.(2) reduces to so called the Wandzura-Wilczek relation.\cite{ww1,ww2}
An interesting question is that how significant is the twist-3 contribution 
in $g_2$. Model predictions (for instance see \cite{song1,jaji}) suggest
that the twist-3 contribution is not small compared to $g_2^{WW}$. 
The earlier data given by E143 Collaboration \cite{e143} and most recent
data given by E155 Collaboration \cite{e155}, however, seem to show that 
$g_2(x,Q^2)$ is close to $g_2^{WW}(x,Q^2)$ and the twist-3 part of $g_2$
is rather small.  
\bigskip

In an earlier unpublished note \cite{song}, we obtained a modified W-W
relation (see Eq.(4) below) with respect to the Nachtmann variable $\xi$.
The result was used by E143 Collaboration \cite{todd}. In this brief 
letter, we present the modified $g_2^{WW}(\xi)$ and compare it with
most recent data. More discussions on kinematical corrections arising 
from the target mass effect are given. 

\bigskip

The modified Wandzura-Wilczek relation is
$${g}_2^{WW}(\xi,Q^2)=-{g}_{1}(\xi,Q^2)+
K_2(\xi,Q^2)\int_{\xi}^{1}{{dy}\over y}\left( 
{{{g_1}(y,Q^2)}\over {K_1(y,Q^2)}}
-6{{M^2y^2}\over {Q^2}}\int_{y}^{1}{{dz}\over {z}}
{{{g_1}(z,Q^2)}\over {K_1'(z,Q^2)}}\right),
\eqno (4)$$
where the kinematic factors $K_{1,2}$ and $K_1'$ are

$$K_1(y,Q^2)={{1-M^2y^2/Q^2}
\over {(1+M^2y^2/Q^2)(1+3M^2y^2/Q^2)}},
\eqno (5a)
$$

$$K_2(\xi,Q^2)={{1-M^2\xi^2/Q^2}\over
{(1+M^2\xi^2/Q^2)^2}},
\eqno(5b)$$

$$K_1'(z,Q^2)={{1-M^2z^2/Q^2}
\over {1+M^2z^2/Q^2}}.
\eqno(5c)$$
The derivation of (4) with (5a-c) is given in the appendix.
Several remarks are in order.

\begin{itemize}

\item{(i) In the large-$Q^2$ limit, all correction factors $K_1(\xi,Q^2)$, 
$K_1'(\xi,Q^2)$ and $K_2(\xi,Q^2)$ given in (5a), (5b), and (5c) 
approach unity and (4) becomes
$${g}_2^{WW}(\xi,Q^2)=-{g}_{1}(\xi,Q^2)+
\int_{\xi}^{1}{{dy}\over y}\left( g_1(y,Q^2)
-6{{M^2y^2}\over {Q^2}}\int_{y}^{1}{{dz}\over {z}}g_1(z,Q^2)\right)
\eqno (6)$$
Considering $M^2/Q^2\rightarrow 0$, and $\xi\rightarrow x$, Eq.(6) 
reduces to the original W-W relation Eq.(3).}

\item{(ii) From Eq.(1), one would expect $\xi_{min}=0$ and 
$\xi_{max}=2/(1+{\sqrt {1+4M^2/Q^2}}~)$ for $x=0\rightarrow 1$. However,
since the true momentum fraction carried by quarks is $\xi$ (if quark is
massless) rather than $x$, hence we should take $\xi_{max}=1$ from the 
beginning. We note that the derivation of Eq.(4) does not depend on the
value of $\xi_{max}$.}

\item{(iii) To show the correction effect, we first plot $K_1$, $K_2$ and
$K_1'$
as functions of $\xi$ for $Q^2=3$ (GeV)$^2$, in Fig.1. One can see that 
for $\xi=0\rightarrow 1$, $K_1=1 \rightarrow 0.324$, $K_2= 1\rightarrow 
0.460$ and $K_1'= 1\rightarrow 0.581$. It seems that all correction 
factors reach the maximum at $\xi=0$. However, their combined effect 
presented in Eq.(4) is not so simple. }

\item{(iv) To show some aspects of the correction effect, we assume that 
$g_1$ in the integral in Eq.(4) is a constant and define a ratio
$$R(\xi,Q^2)=I(K_1,K_2,\xi,Q^2)/I(1,1,\xi,Q^2),
\eqno (7a)$$
where
$$I(K_1,K_2,\xi,Q^2)\equiv K_2(\xi,Q^2)\int_{\xi}^{1}{{dy}\over y} 
\left({1\over {K_1(y,Q^2)}}-6{{M^2y^2}\over {Q^2}}\int_{y}^{1}{{dz}\over {z}}
{1\over {K_1'(z,Q^2)}}\right).
\eqno (7b)$$
The ratios $R(\xi,Q^2)$ for $Q^2=3$, 5, 10 and 100 (GeV/c)$^2$
as functions of $\xi$ are shown in Fig.2. One can see that the ratio is 
quite large at low $Q^2$ and approaches unity when $Q^2\rightarrow\infty$.
However, the ratio (7a) only provides an incomplete 
information of the kinematical target mass correction to $g_2^{WW}$. 
First, the function $g_1(y,Q^2)$ is not a constant but function of $y$, 
and secondly, one should take the whole result from Eq.(4), not just the
second term.}

\item{(v) As pointed out in \cite{ralston} that the original 
Wandzura-Wilczek relation was derived from the Dirac equation for free 
massless quarks and no higher twist corrections were included. By using 
the equation of motion with nonzero quark mass and imposing the gauge 
invariance, an improved Wandzura-Wilczek relation is obtained in
\cite{ralston}
$$g_2(x)=-g_1(x)+\int_x^1{{dy}\over y}g_1(y)
-{{m_q}\over M}\int_x^1{{dy}\over y}
{{\partial h_T(y)}\over {\partial y}}
-\int_x^1{{dy}\over y}\Gamma (y),  
\eqno (8)$$
where $h_T(x)$ is the transverse polarization density and 
$\Gamma(y)$ is related to the multiparton distribution $h_T(x,x')$. 
The quark mass-dependent term ($\sim m_q/M$) in (8) is another twist-2 
piece in addition to the usual term $g_2^{WW}(x)$. The last term in (8) 
is a twist-3 term which is coming from the quark gluon interactions. 
Assuming the $m_q/M$ term and twist-3 contribution are small, we expect 
that a modified version of Eq.(8) with the kinematical target mass
corrections would be very similar to Eq.(4).}

\item{(vi) Making use of the phenomenologically fitted function to the
$g_1$ data, the modified $g_2^{WW}(\xi)$ in Eq.(4), $g_2^{WW}(\xi)$ in
Eq.(6), and the original $g_2^{WW}(x)$ in Eq.(3) are plotted as functions 
of $\xi$ in Fig.3. The data of $g_2(x)$ are taken from E143 \cite{e143} 
and E155 \cite{e155}. From Fig.3, one can see that the effect of the
kinematical target mass corrections is rather small relative to the 
experimental errors. All three $g_2^{WW}$ curves seems to be consistent
with the $g_2$ data. However, more precise data are needed for a 
significant comparison of $g_2^{WW}$ and $g_2$. Most recently, two 
papers \cite{b-t99,p-r98} published on the same topic - target mass 
corrections on the Wandzura and Wilczek relation - which found that 
target mass corrections do not affect the W-W relation (2) if all powers 
in $M^2/Q^2$ are included. We do not know, however, if this conclusion
holds for relation (8). Anyway, since our result (4) holds up to the first
power of $M^2/Q^2$, and target mass corrections have very small effect 
on the W-W relation, hence there is no contradiction between ours and
theirs. }

\item{(vii) It is easy to verify that by changing variable $\xi$ to $x$
and defining $a(x,Q^2)\equiv{\sqrt {1+4M^2x^2/Q^2}}-1$, Eq.(4) can be
rewritten as
$$g_2^{WW}(x,Q^2)=-g_1(x,Q^2)+{{1+a(x,Q^2)/2}\over {(1+a(x,Q^2))^2}}
\int_x^1{{dy}\over{y}}\lbrack {{1+2a(y,Q^2)}\over {1+a(y,Q^2)/2}}
g_1(y,Q^2) $$
$$\qquad\qquad\quad -{{y^2}\over{(1+a(y))(1+a(y,Q^2)/2)^2}}
\int_y^1{{dz}\over {z^3}}3a(z)(1+a(z)/2)g_1(z,Q^2) \rbrack .
\eqno (9)$$
This is the result obtained in \cite{linda}. }

\end{itemize}
\bigskip

\leftline{\bf Acknowledgments}

I would like to thank P. K. Kabir for helpful comments. The author 
thanks L. Stuart for giving Eq.(9) and T. Averett for providing the 
earlier E143 data when \cite{song} was completed. The author thanks 
J. Bluemlein and G. Ridolfi for their useful comments. This work was  
supported by the Institute of Nuclear and Particle Physics, Department 
of Physics, University of Virginia, and the Commonwealth of Virginia. 

\bigskip


\bigskip

\leftline{\bf Appendix}
\bigskip

From Eqs. (47) and (48) in Wandzura's paper \cite{ww2}, for $n$=2,4,...,
we have
$$\int_0^{\xi_{max}}d\xi\cdot \xi^n(1+{{M^2\xi^2}\over {Q^2}})
\lbrack {n\over {n+1}}(1+{{n+2}\over {n+3}}\epsilon){g_1}
+ (1+\epsilon+{{\epsilon^2}\over {n+4}}){g}_2^{WW}\rbrack =0,
\eqno ({\rm I.1})$$
where 
$$\epsilon\equiv\epsilon (\xi,Q^2)\equiv ({{2M^2\xi^2}\over {Q^2}})/
({1-{{M^2\xi^2}\over {Q^2}}}).$$
In obtaining Eq.(I.1), $g_2$ has been replaced by $g_2^{WW}$, or
equivalently, $\bar g_2$ has been neglected.

In the large-$Q^2$ limit, $\epsilon\rightarrow 0$, (I.1) becomes
$$\int_0^{1}d\xi\cdot \xi^n
\lbrack {n\over {n+1}}{{g}_{1}(\xi,Q^2)}
+ {g}_{2}^{WW}(\xi,Q^2)\rbrack=0\quad 
({\rm large}-Q^2\  {\rm limit}).
\eqno ({\rm I.2a})$$
From (I.2a), one easily obtains 
$${g}_2^{WW}(\xi,Q^2)=-{g}_{1}(\xi,Q^2)+
\int_{\xi}^1{{dy}\over y}{g}_1(y,Q^2)\quad 
({\rm large}-Q^2\  {\rm limit}),
\eqno ({\rm I.2b})$$
which is the same form as Eq.(3), but with respect to the variable $\xi$. 
Considering $\xi\rightarrow x$ in the large-$Q^2$ limit, the relation
(I.2b) approaches the original Wandzura-Wilczek relation Eq.(3).

On the other hand, in the large-$n$ limit, (I.1) becomes
$$\int_0^{\xi_{max}}d\xi\cdot \xi^n(1+{{M^2\xi^2}\over {Q^2}})
(1+\epsilon(\xi,Q^2))\lbrack  {g_1}(\xi,Q^2)
+ {g}_2^{WW}(\xi,Q^2)\rbrack =0\quad ({\rm large}-n\ 
{\rm limit}).
\eqno ({\rm I.3a})$$
The main contribution to the integral comes from the large-$\xi$
region due to the suppression factor $\xi^n$. It implies that
$${g}_2^{WW}(\xi,Q^2)\simeq -{g}_1(\xi,Q^2)
\qquad (\xi\rightarrow \xi_{max}).
\eqno ({\rm I.3b})$$
Considering the large-$Q^2$ limit (I.2b) and large-$n$ limit (I.3b), 
it is naturally to assume 
$${g}_2^{WW}(\xi,Q^2)=-{g}_{1}(\xi,Q^2)+K_2(\xi,Q^2)
\int_{\xi}^{\xi_{max}}{dy}f(y,Q^2)\quad ({\rm finite}\ Q^2),
\eqno ({\rm I.4})$$
where $K_2(\xi,Q^2)$ and $f(y,Q^2)$ are two unknown functions to be 
determined and they have the following behavior in the large-$Q^2$ limit 
$$K_2(\xi,Q^2)\rightarrow 1,\qquad f(y,Q^2)\rightarrow 
{{{g}_{1}(y,Q^2)}\over {y}}.
\eqno ({\rm I.5})$$ 

Contrast $K_2(\xi,Q^2)$ with $f(y,Q^2)$, the former is a pure
kinematical correction factor and does not depend on $g_1$. 
Substituting (I.4) into (I.1) and neglecting the $\epsilon^2$ term, 
we obtain
$$\int_0^{\xi_{max}}d\xi\cdot \xi^n(1+{{M^2\xi^2}\over {Q^2}})
[(1+{{2n+3}\over {n+3}}\epsilon){{-g_1}\over {n+1}}
+ (1+\epsilon)K_2(\xi,Q^2)\int_{\xi}^{\xi_{max}}{dy}f(y,Q^2)]=0.
\eqno ({\rm I.6})$$
Unlike the derivation of (I.2b) from (I.2a), we have to use {\it one}
equation, (I.6), to determine {\it two} unknown functions. Since $K_2$ 
is a pure kinematical correction factor satisfies the large-$Q^2$ behavior
(I.5), we may choose 
$$K_2(\xi,Q^2)\equiv (1+\epsilon (\xi,Q^2))^{-1}
(1+{{M^2\xi^2}\over {Q^2}})^{-1},
\eqno ({\rm I.7})$$
and rewrite (I.6) as 
$$\int_0^{\xi_{max}}{{d\xi}\over {n+1}}\cdot \xi^{n+1} 
\lbrack f(\xi,Q^2)-(1+{{M^2\xi^2}\over {Q^2}})
(1+{{2n+3}\over {n+3}}\epsilon){{{g_1}(\xi,Q^2)}
\over {\xi}}\rbrack =0,
\eqno ({\rm I.8})$$
where we have exchanged the order of the integrals in the second term 
of Eq.(I.6).

To determine the second unknown function $f(\xi,Q^2)$, we decompose it
into two pieces 
$$f(\xi,Q^2)=f^{(0)}(\xi,Q^2)+f^{(1)}(\xi,Q^2),
\eqno ({\rm I.9})$$
where $f^{(0)}(\xi,Q^2)\sim O(1)$ and $f^{(1)}(\xi,Q^2)\sim
O(M^2/Q^2)$ is a small term. Since $f^{(1)}(\xi,Q^2)\rightarrow 0$ in 
the limit $Q^2\rightarrow\infty$, the function $f^{(0)}(\xi,Q^2)$ must
satisfies large-$Q^2$ behavior (I.5). We purposely choose 
$$f^{(0)}(\xi,Q^2)=(1+{{M^2\xi^2}\over {Q^2}})
(1+2\epsilon){{{g_1}(\xi,Q^2)}\over {\xi}}.
\eqno ({\rm I.10})$$
From (I.8), (I.9), and (I.10), we have
$$\int_0^{\xi_{max}}{{d\xi}\over {n+1}}\cdot \xi^{n+1}
\lbrack f^{(1)}(\xi,Q^2)+(1+{{M^2\xi^2}\over {Q^2}})
{{3\epsilon(\xi,Q^2)}\over {n+3}}{{{g}_1(\xi,Q^2)}
\over {\xi}}\rbrack=0.
\eqno ({\rm I.11})$$

To determine small unknown function $f^{(1)}(\xi,Q^2)$, we put
$$f^{(1)}(\xi,Q^2)\equiv \xi~\int_{\xi}^{\xi_{max}}dy\eta(y,Q^2),
\eqno ({\rm I.12})$$
where $\eta(y,Q^2)$ should be the order of $O(M^2/Q^2)$. 
Substituting (I.12) into (I.11), one obtains
$$\int_0^{\xi_{max}}d\xi{{\xi^{n+3}}\over {(n+1)(n+3)}}
\lbrack \eta(\xi,Q^2)+3\epsilon(\xi,Q^2)(1+{{M^2\xi^2}\over {Q^2}})
{{{g_1}(\xi,Q^2)}\over {\xi^3}}\rbrack=0.
\eqno ({\rm I.13})$$
This equation can be satisfied for all $n$ ($n$=2,4,...) only if 
the term in the bracket vanishes. Therefore
$$\eta(\xi,Q^2)=-3\epsilon(\xi,Q^2)(1+{{M^2\xi^2}\over {Q^2}})
{{{g_1}(\xi,Q^2)}\over {\xi^3}},
\eqno ({\rm I.14})$$
which is indeed the order of $O(M^2/Q^2)$. Substituting (I.9), (I.10), 
(I.12), and (I.14) into (I.4), we finally obtain the modified W-W relation 
Eq.(4) with correction factors (5a), (5b), and (5c).

\begin{figure}[h]
\epsfxsize=6.0in
\centerline{\epsfbox{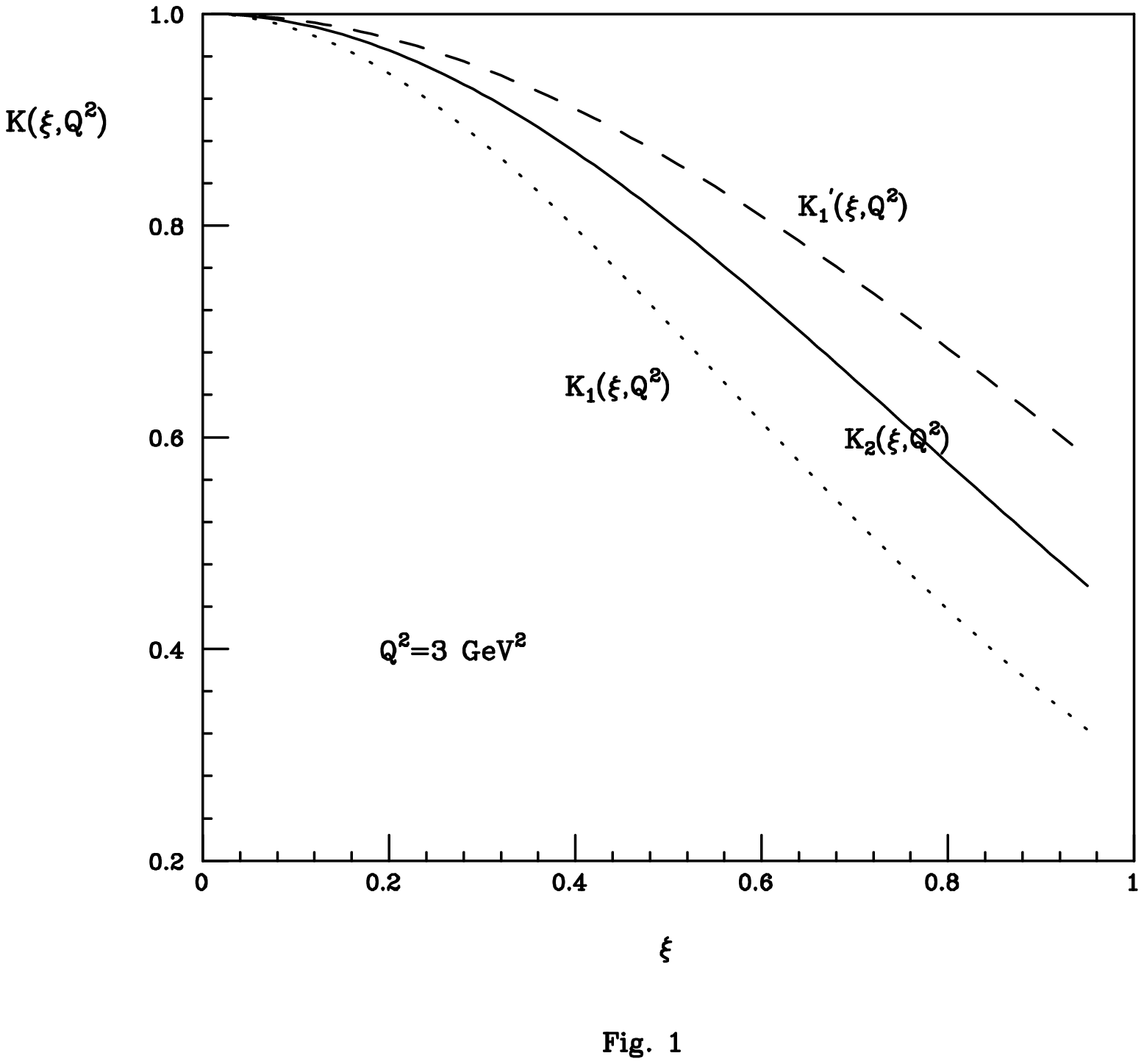}}
\caption{The kinematical correction factors (see Eqs.(5a), (5b), and (5c))
plotted as functions of the Nachtmann variable $\xi$ at Q$^2$=3
(GeV/c)$^2$.}
\end{figure}

\begin{figure}[h]
\epsfxsize=6.0in
\centerline{\epsfbox{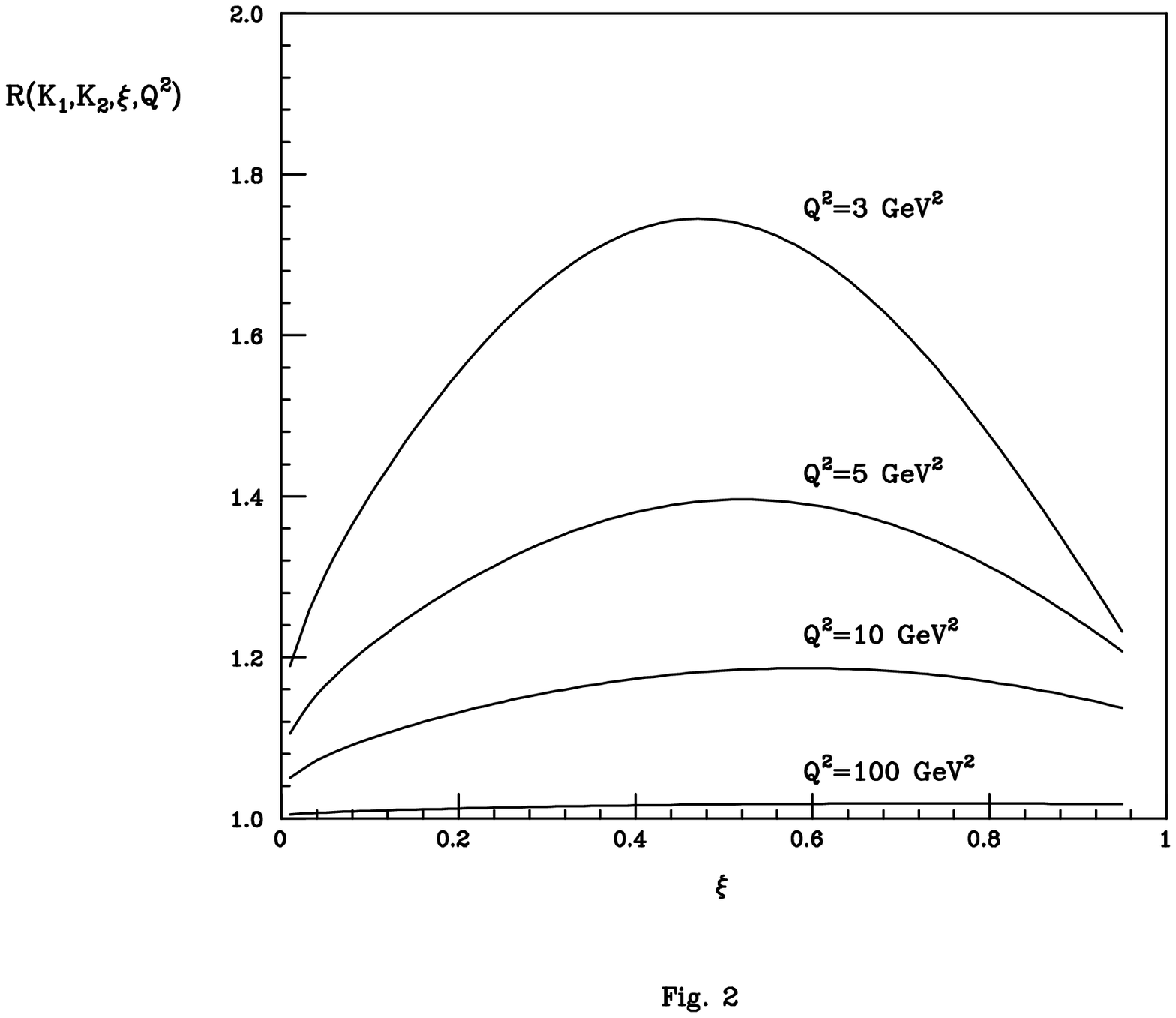}}
\caption{The correction ratios (see Eq.(6a,b)), plotted as functions 
$\xi$, for Q$^2$=3, 5, 10 and 100 (GeV/c)$^2$.}
\end{figure}

\begin{figure}[h]
\epsfxsize=6.0in
\centerline{\epsfbox{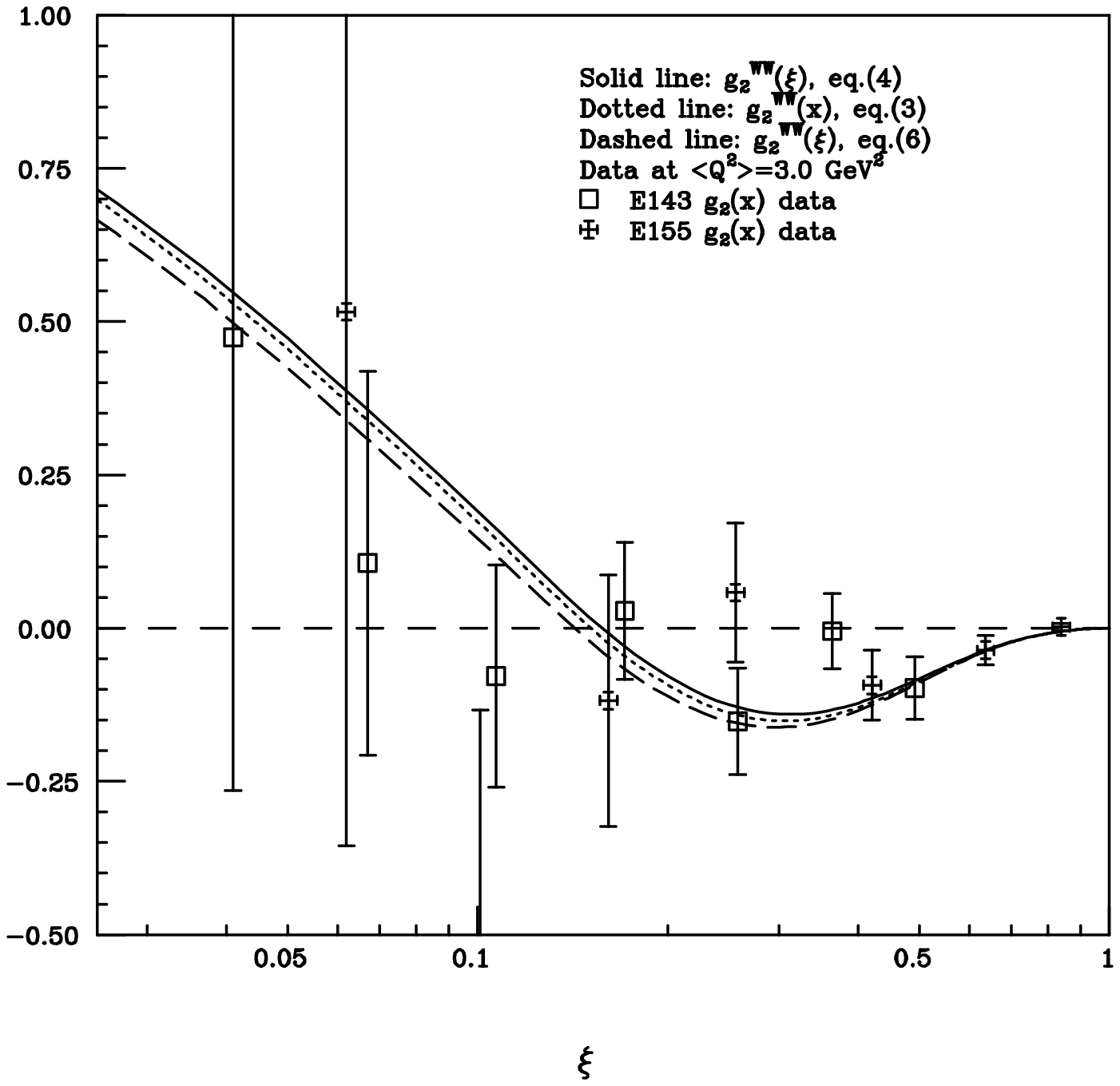}}
\caption{Modified $g_2^{WW}(\xi)$, Eq.(4) and Eq.(6), and the original 
W-W relation, Eq.(3) plotted as functions $\xi$, for Q$^2$=3 (GeV/c)$^2$.
Data are taken from E143 and E155.}
\end{figure}

\end{document}